\documentclass[a4paper]{mem}
\usepackage{natbib}
\usepackage{graphicx}
\newcommand{\chandra}{\textit{Chandra}}
\newcommand{\einstein}{\textit{Einstein}}
\newcommand{\xmm}{\textit{XMM-Newton}}
\newcommand{\hst}{\textit{HST}}

\newcommand{\rosat}{\textit{ROSAT}}
\begin{document}
   \title{A high-energy view of radio-loud AGN}

   \author{D.M. Worrall}


   \institute{Department of Physics, University of Bristol, Tyndall
Avenue, Bristol, BS8 1TL}

   \abstract{
Seyfert galaxies and quasars were first discovered through optical and
radio techniques, but in recent years high-energy emission, that can
penetrate central gas and dust, has become essentially the defining
characteristic of an AGN.  AGNs with extended radio jets are of
particular interest, since the jets signal source orientation.
However, the jets extend into the cores, where they are faster and
more compact. Special-relativistic effects then cause jet brightness
and variability time-scales across the electromagnetic spectrum to be
strong functions of jet orientation.  Jet X-ray emission is confused,
to varying degrees, with that from the central engine, but can be
measured, at least in a statistical sense, through considerations of
the multiwaveband spectrum and the level of intrinsic absorption.  The
rich high-energy structures found in jets which are resolved with
\chandra\ and \hst\ inform our interpretation of the inner structures.
In particular, it is found that shocks are prevalent and don't
necessarily disrupt jets, and that one-zone models of emission near
shocks are an over-simplification.
   }
   \authorrunning{D.M. Worrall}
   \titlerunning{A high-energy view of radio-loud AGN}
   \maketitle
%

\section{Introduction}

Now that it is known that massive or supermassive black holes are
prevalent, if not ubiquitous, at the centres of galaxies, the line
between active and non-active galactic nuclei has become harder to
define.  We must ensure that the perceived level of activity is not
underestimated because radiation is obscured by gas and dust.  Indeed
it is often the case, and a prediction of unified schemes, that the
central emission of an active galactic nucleus (AGN) must penetrate
large columns of gas and dust.  High-energy emission, since it is able
to penetrate dust, and at the highest X-ray energies to penetrate gas,
is now virtually the defining characteristic of an AGN.  But the
high-energy emission is complicated and there remain unanswered
questions.  Active galaxies with radio jets are arguably the most
complicated AGN. To ascertain if and how the central engines differ
from non-jetted AGN, we need to peel off the jet emission to reveal
the central engines.

Here the \chandra\ X-ray Observatory \citep{weis00} has come into its
own.  The jet components it has resolved have informed our
interpretation of the unresolved structures and allowed much
interesting and important physics to be addressed.  Since a complete
discussion of the high-energy emission of radio-loud AGN is beyond the
scope of this review, I'll concentrate mostly on one aspect,
the extended components, since this has been a particular area of
intense interest and study over the last few years.

\section{Foundations}

Back in the late 1970s, when the \einstein\ Observatory found a
typical AGN to be a bright X-ray source, there were two important
discoveries.  Both were based on the total observed X-ray emission.
Firstly, it was found that radio-loud quasars were brighter X-ray
sources than radio-quiet quasars for a given optical luminosity
\citep{ku80, zamorani81}.  The dependence of X-ray luminosity and
spectrum on beaming angle that was subsequently found was interpreted
as evidence that the X-ray emission was composed of radio-quiet (i.e.,
accretion-flow related) and radio-loud components \citep{zamorani86,
browne87, worrall87, wilkes87, canizares89, kembhavi93, shastri93}.
It was only speculation that the component related to the accretion
flow might be the same in radio-quiet and radio-loud quasars, even
though the quasar nature implied that the sources were all radiating
with high Eddington accretion efficiency.

Secondly, a correlation of X-ray emission with core radio emission was
found for radio galaxies \citep{fabb84}, i.e., sources presumed to be
non-beamed counterparts of quasars and BL~Lac objects.  This was
supported by \rosat\ observations that largely separated the core
X-ray emission from the thermal emission of the host galaxies and
clusters \citep{wb94, canosa99}.  The implications were either (a)
that the central-engine X-rays were correlated with radio loudness or
(b) that the X-rays were from an inner radio jet, and the
central-engine X-rays were obscured or weak.  In the 1990s, after our
first such correlations using \rosat, I pointed out \citep{wcrac} that
the brightest features in the X-ray-resolved jets of the nearest radio
galaxies, Cen~A and M~87, fitted on the same correlation as the core
emission from more distant radio galaxies.  This gave support to
interpretation (b), and implied that \chandra\, with its sub-arcsecond
spatial resolution, should resolve jets in radio galaxies more distant
than Cen~A and M~87.

\section{Resolved jets in the \chandra\ era}

\subsection{Low-power radio sources}

Low-redshift sources provide the best linear spatial resolution, so it
is interesting to look at them first.  Naturally, these sources are
biased towards having low radio power.  Here resolved jet emission is
seen in sources of all orientation, from two sided jets in 3C~270
\citep[e.g.,][]{chiab4261, zezas04}, where the jet emission is
relatively faint and hard to see in contrast with X-ray emission from
the host galaxy NGC~4261, to the stubby one-sided jets of BL Lac
objects \citep[e.g.,][]{pesce01, birkins02} where it is believed that
the jets are at small angle to the line of sight and we are seeing
boosted emission from the relativistic jet pointed towards the
observer.  Typically jets are shorter in the X-ray than the radio.
The X-rays are often associated with regions in which the jets are
believed to be decelerating through entrainment of the ambient
X-ray-emitting medium \citep[e.g.,][]{hard31}.  It is natural to
consider shocks associated with this entrainment as being responsible
for the knotty structures of many of the jets.

There are selection effects in the \chandra-observed sources, with a
bias in favour of those where the kpc-scale radio emission is
predominantly one sided, and known optical jet sources are
over-represented.  It has been normal to detect one-sided jet X-ray
emission, and indeed there is better contrast with galaxy emission in
the X-ray than in the optical to the depth of \hst\ snapshot surveys
\citep{worrlowz}.  A broken power law normally fits the radio, optical
and X-ray spectrum, integrated over the entire X-ray-emitting region
\citep[e.g.,][]{hard66, boeh01, birkins02}.
Typically model predictions for inverse Compton (iC) emission in the
X-ray band fall short of the observations \citep[e.g.,][]{hard66} and
the X-ray spectrum is, in any case, too steep to be iC from the
electrons producing the radio synchrotron emission
\citep[e.g.,][]{boeh01}.

A synchrotron frequency can be related to a most probable energy for
the emitting electron, given an adopted magnetic field strength, $B$.
It is normal to adopt the value for $B$ for which the source contains
a minimum total energy in particles and magnetic field.  Such
minimum-energy values for $B$ have been confirmed to within a factor
of a few in most hotspots and radio lobes observed with \chandra\ and
\xmm, where $B$ is measured through associating the X-ray emission
with inverse Compton scattering of the population of electrons that is
producing the radio synchrotron emission \citep[e.g.,][]{hardhot,
brun02, isobe02, comastri03, bondi04, belsole04, crost04}.
Minimum-energy values of $B$ in low-power jets are typically around
100~$\mu$G.  X-ray-emitting electrons then lose their energy in tens
of years, which is less than the plasma travel time from the core,
implying the need for {\it in situ\/} acceleration.  It is natural to
invoke the existence of shocks.  The shocks presumably have a range of
strengths, and the fact that the multiwavelength spectra and images
are integrals over inhomogeneous regions may explain two observed
features.  Firstly, the size of the break in the spatially integrated
multiwavelength spectrum is too large by $\Delta\alpha \sim 0.2$ for a
simple continuous injection and energy-loss model
\citep[e.g.,][]{birkins02}.  Secondly, in sub-regions there are often
offsets between peaks of X-ray and radio emission, with the X-ray
typically lying closer to the core \citep[e.g.,][]{hard66, hardcena}.

The greatest detail is seen in the closest radio galaxy, Cen~A.  In
the X-ray the jet is intrinsically blobby, with an apparent filling
factor of less than about 10 per cent.  Here it has been proposed
\citep{hardcena} that, in at least some locations, shocks are the
result of obstacles in the jet (gas clouds or high-mass-loss stars)
such that both radio-emitting and X-ray-emitting electrons are
accelerated in the standing shock ahead of the obstacle, and
downstream a wake produces further acceleration of just the low-energy
electrons that emit in the radio.  The sort of radio-X-ray offset that
this model describes, averaged over several knots, might explain the
radio-X-ray offsets seen in more distant jets.

Polarization fraction and direction changes are a signature of shocks.
Optical frequencies are particularly good for probing this, since
Faraday rotation is negligible.  In M~87 there is evidence for strong
shock acceleration at the base of bright emitting regions in
compressed transverse magnetic fields \citep{perl99}.  Multiwavelength
variability studies provide another probe of acceleration and energy
losses.  Here again it is M~87 that has so far proved the best
resolved jet for study in this way \citep{harris03, perl03}.

The synchrotron jets in low-power sources thus provide test-beds for
studies of particle acceleration.  It appears that shocks are
accelerating particles and changing magnetic field directions in
regions where the jets, although slowing, can still be propagating at
speeds more than a few tenths of the speed of light and are still
well collimated.  If this behaviour applies at higher speeds,
it suggests that one-zone models should always be treated with
caution, even imbedded in small-scale blazar jets.

\subsection{High-power radio sources}

High-power jets are rarer and so more distant.  The beamed
counterparts are the quasars.  Observing them was not initially a high
priority for \chandra\ since it was recognized that the cores were
bright, and the likelihood of multiple photons arriving between CCD
readouts was high, leading to distorted spectral measurements (so
called `pileup').  It was fortuitous that a radio-loud quasar was the
chosen target for in-flight focus calibration, since this led to the
detection of resolved jet emission from the $z = 0.651$ quasar PKS
0637-752 \citep{schwartz00, chartas00}.  Major programs targeting the
jets of core-dominated quasars followed \citep{sambruna02, sambruna04,
marshall04}, with a detection success rate of roughly 50 per cent.

The generally favoured X-ray energy-production mechanism for the
quasar jets has been inverse Compton (iC) scattering of cosmic
microwave background (CMB) photons by a fast jet (with a
minimum-energy magnetic field) which sees boosted CMB and emits beamed
X-rays in the observer's frame \citep{tavecc00, celott01}.  The
mechanism requires the jets to have a highly-relativistic bulk flow
and be at a small angle to the line of sight, as expected for
core-dominated quasars. Although in PKS 0637-752 such a speed and
angle are supported on the small scale by VLBI measurements
\citep{lovell00}, the fast speed must persist up to hundreds of kpc
from the core (after projection is taken into account) for the X-rays
to be produced by this mechanism.  Single-zone synchrotron
self-Compton (SSC) models lead to an uncomfortably large departure
from minimum energy (a factor of about 1000 in the case of PKS
0637-752). Optical emission falling below a spectral interpolation
between the radio and X-ray has been used to rule out synchrotron
radiation from a single population of electrons as the explanation of
the spectral energy distribution for this source \citep{schwartz00}.

Clearly as long as jets remain fast out to hundreds of kpc from the
core, the beamed iC-CMB mechanism produces X-ray emission at some
level.  Since the CMB energy density increases as $(1 + z)^4$, the
surface brightness for this mechanism is constant with redshift, while
the radio-synchrotron surface brightness decreases with redshift.  But
there is a difficulty with the beamed iC-CMB interpretation.  Sharp
gradients in X-ray surface brightness (sharper than in the radio) at
the edge of knots \citep[e.g.,][]{chartas00} or X-ray emission
decreasing with distance along the jet while the radio increases
\citep[e.g.,][]{sambr273, sambruna04, marsh273}, sometimes with
distinct radio-X-ray offsets \citep[e.g.,][]{siem02, jm04} are not
naturally explained, since with the beamed iC-CMB mechanism the X-rays
are from low-energy electrons with long energy-loss lifetimes.  This
may suggest that jets are clumpy \citep{tavecc03}, but in this case
the SSC may be enhanced and the requirement for fast jets and the
importance of beamed iC-CMB may diminish \citep{schwartz00}.  It may
suggest jet decelerations, with declining effectiveness of the iC-CMB
process along the jet accompanied perhaps by magnetic-field
compression that produces more radio synchrotron emission
\citep{sambr273, gk04}.  But, it remains possible that the X-rays are
dominated by synchrotron emission, either from high-energy electrons
whose efficiency in losing energy by inverse Compton scattering is
decreased through being in the regime where the Klein-Nishina cross
section applies \citep{da02}, or from a separate electron population,
perhaps due to transverse velocity structure in the jet \citep{jm04}.

To assess the importance of X-ray synchrotron emission in the resolved
jets of powerful radio sources, it is beneficial to select some of the
nearest powerful radio galaxies, where the jets should not be at the
same small angles to the line of sight deduced for core-dominated
quasars, thus de-emphasizing beamed iC-CMB emission.  Good examples of
such radio galaxies for which the case is made for X-ray synchrotron
emission are 3C 403 \citep{kraft04} and 3C 346 \citep{wb346}.  3C 346
is of particular interest because offsets between radio and X-ray
emission are seen in a bright knot where the jet makes a dramatic
change in direction.  The observations have been interpreted as due to
an oblique shock in the wake of a companion galaxy's track though the
host cluster, and predictions are made for a change in magnetic field
direction that are testable with \hst\ polarimetry.

There remains much to be learned about the X-ray emission mechanisms of
powerful jets, and the relative importance of shocks in accelerating
particles and shaping the morphologies.  An extreme source which may
advance understanding greatly though an upcoming programme of more
sensitive multiwavelength measurements is PKS 1421-490 (Gelbord et
al. 2005, in preparation).  It has an undistinguished radio knot which
is remarkably bright in the optical and X-ray, with multiwavelength
properties that challenge straightforward emission models.

\section{X-ray Cores}

In highly boosted sources (e.g., core-dominated quasars and BL Lac
objects) it is common to assume that jet emission dominates the
continuum at all energies, including the X-ray, where it is
assumed to swamp emission associated with the accretion flow (see
section 2).  One-zone SSC models, often including external photon
fields, are commonly applied, typically implying emission regions of
order $10^{16}$~cm in size and magnetic fields of order a Gauss
\citep[e.g.,][]{gis98, tag03}.  Sometimes correlated multiwavelength
flares support the presence of a dominant emission region
\citep[e.g.,][]{urry97, tak00}, but in other cases uncertainties of
size scales, geometries, and parameters for the competing processes of
energy loss and acceleration force adoption of poorly-constrained
models.  On resolved scales we find evidence that shocks don't
necessarily disrupt a fast flow and multiple synchrotron-emitting
emission regions appear.  This may suggest that one-zone models are an
oversimplification for the inner jets.

The cores of non-boosted powerful radio galaxies are more difficult to
study in the X-ray.  Distance, and the obscuring torus invoked by
unified models that should weaken their nuclear flux at low X-ray
energies, gives \xmm\ advantages over \chandra\ for their study.
However, kpc-scale jet emission then appears in the unresolved cores
\citep[e.g.,][]{belsole04}, complicating the issue of component
separation.

Low-power sources are closer and the cores are more easily X-ray
studied \citep[e.g.,][]{donato04}, but ideas are divided about the
origin of this emission.  Using NGC 6251 as an example, one school
looks at spectral energy distributions and interprets the emission in
terms of synchrotron and iC jet models \citep[e.g.,][]{chiab6251}.
The other emphasizes the detection of variability or Fe-K line
emission and models the sources in terms of an accretion disk
\citep[e.g.,][]{gliozzi04}.  However, there are least two sources
which show spectral evidence for both components.  The first
\citep{evans04} is Cen~A.  As our closest radio galaxy, the X-ray jet
emission is better resolved out from the core than for any other
source.  There is strong Fe-line emission which the \chandra\ gratings
resolve but find to be sufficiently narrow that it can be located at
an obscuring torus, with $N_{\rm H} \sim 10^{23}$ cm$^{-2}$,
responsible for absorbing the strong inner emission.  Additional,
somewhat less absorbed, X-ray emission can be associated with the
inner radio jet.  The second source is 3C~270 \citep{zezas04}, which
is currently the best example of a two-sided X-ray jet source. The
X-ray spectrum shows a strong contribution at low energies from
unresolved galaxy gas, but a good model fit contains also Fe-K line
emission and two power-law components, the more absorbed of which can
be associated with the accretion flow and the other with the inner
jets \citep[for alternative interpretations of a single nonthermal
component in this source see][]{chiab4261, sambr4261, gliozzi03}.  In
these sources, at the lowest X-ray energies the dominant component is
interpreted by \citet{evans04} and \citet{zezas04} as associated with
the inner radio jets.  Since \rosat\ was sensitive only to low-energy
X-rays, this can explain the radio and X-ray correlations found by
that satellite and discussed in section 2.

A new approach to studying the cores is to assume that they do indeed
all host a moderately large obscuring torus of $N_{\rm H} \sim
10^{23}$ cm$^{-2}$, and find the upper limit to the luminosity of
X-ray emission behind this column (Evans et al. 2004, in preparation).
A typical value is $10^{41}$ ergs s$^{-1}$, which is similar to the
luminosities of the obscured components measured in Cen~A and 3C~270.
Of course, if even higher obscuring columns are present, these cores
could contain even more luminous components, and the maximum
luminosity and column density are linked by the assumed structure of
the obscuring torus.

\section{Summary}

This review has concentrated on the X-ray observations of the jets and
cores of AGN, the emission mechanisms that have been ascribed to
them, and the observed morphologies. 

The resolved X-ray jets of low-power radio sources can be interpreted
as synchrotron radiation in regions where the jets, although
slowing by entrainment, are still believed to travel at significant
bulk speed.  Short electron energy-loss times
then require {\it in situ} particle acceleration that is most naturally
attributed to the presence of shocks.  Although much detail is still to
be explored and understood, such a picture can qualitatively explain
observed spectral breaks in the radio, optical, and X-ray spectrum, and
spatial offsets in X-ray and radio emission regions.

The origin of the X-rays in the resolved jets of high-power sources is
more controversial.  Synchrotron emission dominates in some cases, and
again shocks may be important in particle acceleration.  The situation
is less clear in sources at high redshift, and where the jets are
close to the line of sight.  Here inverse Compton processes are
expected to be more important, and may dominate.

What is defined as the core emission region depends on instrument
resolution and source redshift.  It is natural that some of this X-ray
emission should be associated with the radio jets, and what is learned
from the resolved jets can inform our interpretation of this
component.  What is less clear in many sources is how much X-ray
emission is associated with the accretion flow, and whether or not a
central structure obscures much central emission.

Over the past decade, it is not just the jet \emph{emission mechanisms} that
have been studied.  Much has been learned about other physics of the
resolved components, particularly through relating properties of the
jets to those of the X-ray-emitting medium through which they
propagate, and which provides sufficient pressure to be dynamically
important.  A discussion of this physics is out of the scope of this
paper, but see \citet{wbrev} for a recent review and discussion of
outstanding issues.

\begin{acknowledgements}
     I am indebted to collaborators, postdocs and students for
     their contributions to the advancement of  understanding and knowledge
     of this topic.
     In particular I wish to thank Mark Birkinshaw
     and Martin Hardcastle.  I thank PPARC for
     ongoing financial support for a postdoctoral researcher, and
     the organizers of the meeting for travel support.
\end{acknowledgements}

\bibliographystyle{aa}

\end{document}